# Filling the gaps in video transcoder deployment in the cloud

Vibhoothi[1], Daniel Joseph Ringis[1], Xin Shu[1], François Pitié[1], Zsolt Lorincz[2], Philippe Brodeur[2], Anil Kokaram[1]
[1] Trinity College Dublin  [2] Overcast HQ
[1] Sigmedia Group, Department of Electrical Engineering, Trinity College Dublin, Dublin 02, Ireland;
[2] Overcast HQ, Dublin 02, Ireland
{vibhoothi, ringisd, xins, pitief, anil.kokaram}@tcd.ie, {zsolt, philippe}@overcasthq.com

**Abstract** – Cloud based deployment of content production and broadcast workflows has continued to disrupt the industry after the pandemic. The key tools required for unlocking cloud workflows, e.g., transcoding, metadata parsing, streaming playback, are increasingly commoditized. However, as video traffic continues to increase there is a need to consider tools which offer opportunities for further bitrate/quality gains as well as those which facilitate cloud deployment. In this paper we consider pre-processing, rate/distortion optimisation and cloud cost prediction tools which are only just emerging from the research community. These tools are posed as part of the per-clip optimisation approach to transcoding which has been adopted by the large streaming media processing entities but has yet to be made more widely available for the industry.

## Introduction

Video transcoding is a critical engineering component in broadcast systems today. As encoders mature, mechanisms to squeeze even more quality at lower bitrate continue to emerge that exist outside the standardisation effort. Per-clip transcoding or content-aware transcoding, deployed at scale by Netflix in 2015 [1], exploits the idea that adapting transcoder parameters to each clip allows substantial gains over generic parameter sets. This is only one aspect of content adaptive treatment. Taking a wider perspective, we admit consideration of pre-processing tools and, from a customer standpoint (particularly in cloud deployment [2]), tools that can predict the compute load about to be incurred. However, i) per-clip content adaptation provides huge gains in compression but comes with substantial compute cost; ii) pre-processing of media is well known to improve compression behaviour, but the control of that pre-processor is difficult; and iii) users cannot yet assess or predict the costs of these modules in the cloud when applied to their datasets without actually launching the job and making a crude prediction as the job unfolds. In this paper we address these gaps as follows.

1. **Optimised Pre-processing:** The video denoiser is one of the simplest pre-processing tools that has a direct impact on video quality and bitrate. Since the late 1990's [3, 4] the potential was recognised for video denoisers to reduce bitrate in compression and improve picture quality especially for User Generated Content (UGC). But controlling the interaction of the denoiser with the encoder remained a challenge especially if there was no integration of the two processes. Too much denoising drives the bitrate down, but it destroys picture quality. In this paper we introduce a simple, empirical experiment which allows a user to directly measure the interaction and so parameterise the pre-processor optimally for a particular codec/denoiser combination. We show as much as 2dB gains as compared with ignoring this interaction.

2. **Optimised Transcoding:** Discovering those optimal parameter sets is computationally challenging [5]. In the asymmetric world of the large, high-quality streamer, the compute cycles spent in transcoding are expected to be recovered many times over in streaming to the end user. But as a commodity process, per-clip transcoding needs to be managed to control the compute cycles required. This is especially acute with more recent standards like H.265 and AV1 which are notorious for cycle consumption in default configurations. We present in this paper a computationally efficient strategy for optimising the rate distortion criterion in a codec to adapt it to a particular piece of content. The objective gains are measured in terms of Bjøntegaard-Delta (BDRATE) percentage. Gains are modest on average (2% BDRATE) but per-clip gains can hit over 14.89%. Our proxy-based approach can achieve 80% of these gains with over 10x improvement in compute load.

3. **Compute Load Prediction:** Finally, we consider a process for predicting the compute load required to complete a transcoding job. It is true that the problem of predicting "time to complete" is a fundamental challenge in a general software reliability sense. However, in the context of cloud deployment the community is missing a dependable tool for allowing a customer to allocate appropriate budget for a transcoding job, perhaps at scale. Efforts to build such a tool using ML approaches have already been undertaken in the research community [6]. But an examination of datasets shows that it is a challenge to collect enough examples of encode feature/duration pairs so that the feasible space is covered. In this paper, we take a different approach, based on classification and non-linear quantisation of encode durations. Our results show good performance over a wide range of compute loads but still require careful attention to training data and deployment scenarios.

## Section 1: Pre-processing for improved compression performance

Denoising as a pre-processing module has been explored since the 1990's [3, 4]. It is proposed to mitigate the impact of degraded input video on transcoding (e.g., User-Generated-Content (UGC) input content). It is also used to massage input clips so that they become more amenable to encoding and therefore result in less encoding artefacts (e.g., treatment of film grain). The problem is that encoders themselves act as denoisers because of the coring effect of Discrete-Cosine Transform (DCT) quantisation. Van Roosmalen et al [3] were among the first to spot that the effect changed with target bitrate. That means that for effective pre-processing, the denoiser settings should be related not only to the input measured noise level but also to the target bitrate of the encoder. The nature of the encoder itself is also a factor. For example, H.264 will quantise in the DCT space in a different way compared to AV1.

Ideally, this relationship can be implicitly resolved by constraining the denoiser to optimise the quality of the output pictures from the encoder as a function of noise level and target bitrate, and not w.r.t. the quality of the pictures from the denoiser itself. This is an important point. Given an input degraded video sequence $G_n$, we do not care about the output quality of the denoised picture from the pre-processor, i.e., $\hat{I}_n = P(G_n)$ where $P$ is the pre-processor. We only care that the output quality of the denoiser/encoder cascade is high, i.e., $\hat{I}_n^T = T(P(G_n), r)$, where $r$ is the target encoding bitrate and $T()$ is the transcoding operation. The resulting cost function for the denoiser $(I_n - \hat{I}_T)^2$ is therefore highly non-linear because of the influence of the encoder.

Recent efforts using a deep neural network (DNN) as a pre-processor [7, 8, 9, 10] have employed differentiable simulations of the essential parts of an encoder. This offers a fully adaptive module which could adapt to transcoder target bitrate and input noise level. In that case, the DNN functions as much more than a denoiser, acting to enhance details which might be otherwise lost in encoding.

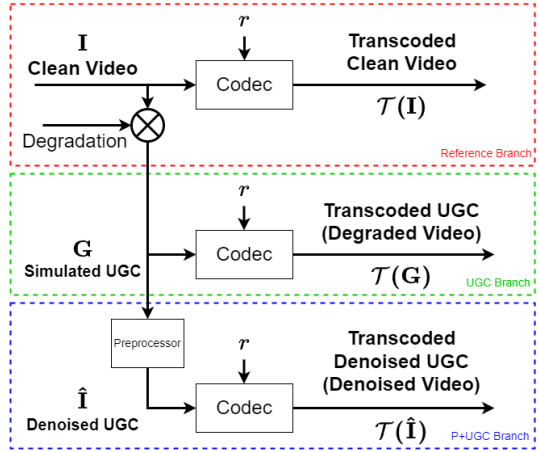

FIGURE 1: WORKFLOW FOR MEASURING OPTIMAL PREPROCESSOR PARAMETERS.

While effective, the computational cost of that module is of course a challenge. In addition, the efficacy of those systems relies on the goodness of fit of the simulated encoder. Instead, we take an empirical approach to the problem. Rather than design a fully adaptable pre-processor, we evaluate its performance using $(I_n - \hat{I}_n^T)^2$, i.e., in tandem with the encoder for a large UGC test set and over all our denoiser parameter settings and target bitrates. This gives us a mapping from input noise level $\sigma^2$ and target bitrate $r$ to the optimal parameter setting for our chosen denoiser and codec combination.

Figure 1 illustrates our experimental setup. A Clean video ($I_n$) is artificially degraded with gaussian noise to create simulated UGC content ($G_n$). This UGC is denoised through the action of a pre-processor resulting in an estimate ($\hat{I}$) of the clean data. At a target bitrate $R$ we require the encoded, preprocessed content output from the P+UGC workflow, to be as close as possible to the reference workflow output. The output of the UGC workflow (middle) is expected to be the worst output content in this workflow model. For the simulations, we use the MCL–JCV (USC MediaComm Lab Set) [11] corpus of 30 clips at 720p in a variety of framerates. We assume the data to be relatively clean from degradation and hence simulate UGC by adding noise to generate PSNR of $G_n$ from 20 to 40 dB. We evaluate both the 3D Wiener filter [12] and an example of a DNN Denoiser fast DVDNet [13] in cascade with both AV1 and VP9 encoders. Further details are available in [14].

Figure 2 shows average PSNR over our dataset of $\hat{I}_n^T$ w.r.t the clean original source using two different preprocessors with AV1, and input degradation level at 27.5 dB. Each curve shows the output quality with varying filter strength at a particular bitrate. In all cases, even with no filtering, the output quality is higher than the input, but there is clearly a different optimal filter strength (corresponding to max PSNR) depending on the target bitrate. The gain in PSNR over no filtering is about 1-2 dB for > 512 Kbps. The denoising effect of the encoder is shown at the left edge of each curve where the filter strength is 0. The curves are not flat but with a clear maximum showing that there is an optimal filter strength for a particular bitrate at this level of input degradation. Gains are almost 2dB over no filtering, for a midrange of bitrates 2-4 Mbps.

The decision surface is smooth enough to use a polynomial for fitting. Given a measured degradation noise level $\sigma$ and the target bitrate $r$ we fit a 5th order polynomial $f(\sigma, r)$ yielding the required optimal filter strength. We employ different models for each codec/filter combination and the resulting prediction reduces the possible PSNR gains by at most 17% for the worst degradation level 20dB. Otherwise, there is little impact of errors in prediction. Figure 3 shows visual results from one filter/codec combination.

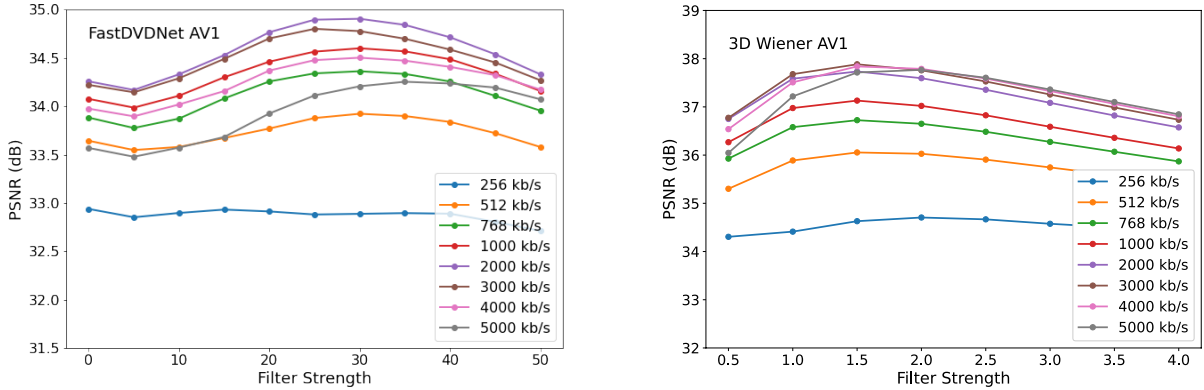

FIGURE 1: PERFORMANCE OF TWO DIFFERENT PRE-PROCESSORS FOR VARYING FILTER SETTINGS AT DIFFERENT BITRATES.

## Section 2: Content-adaptive transcoding for reducing bitrate

Streaming services such as Netflix have relatively small libraries of video for distribution to users at high quality and low bitrate [1]. Aaron et al [1] recognised that by choosing codec parameters per clip, significant gains could be achieved. This is because the statistics of video clips vary greatly over any corpus. Hence, tuning for average improvement does not achieve the best compression performance possible for a given clip. Their work in per clip encoding focused on adjusting input parameters of video transcoding, primarily optimising the bitrate ladder for Dynamic Adaptive Streaming over HTTP (DASH) streaming.

In the work on content-adaptive transcoding, it is typical to adjust parameters like quantisation step size and the resolutions used in the design of the bitrate ladder [15]. Parameters that affect the rate/distortion trade-off are typically left untouched. In this work we investigated per-clip optimisation of the Lagrangian multiplier used in the Rate Distortion Optimisation module of video codecs.

### Rate-distortion Optimisation

Rate-distortion optimisation is a constrained optimisation problem in which the goal of the encoder is to minimise distortion, D, while ensuring that the bitrate, R, remains below some bitrate, $R_{max}$. This problem can be solved using Lagrangian optimisation [16]. In this approach the codec transforms the constrained optimisation problem into an unconstrained problem by introducing a cost function, J, in the rate distortion equation as follows:

$$J = D + \lambda R \qquad (1)$$

This essentially combines distortion and rate through the action of the Lagrangian multiplier ($\lambda$). This technique was adopted in video codecs through empirical experiments starting early as H.263 as it is effective and conceptually simple. However, in practice there are several interactions between coding decisions which make this problem less straight forward, resulting in a number of empirical approximations for transformation of $\lambda$ between different tools in practical encoder implementations. While theory suggests that the selection of $\lambda$ in the encoder should be optimal, in practice it is not optimal for a given clip. There exists a limited amount of work on adaptation of $\lambda$ for this purpose [17]. The idea is typically to adjust lambda away from the codec default by using a constant $k$ as follows:

$$\lambda = k * \lambda_o \qquad (2)$$

where $\lambda_o$ is the default Lagrangian multiplier estimated in the video encoders, and $\lambda$ is the updated Lagrangian. In this work we estimate $k$ per clip by directly maximising BDRATE [18] gains. BD-rate [18] measures the % difference in area between two Rate-Distortion (RD) curves and hence quantifies the improvement in encoding over a range of Quantization-parameters (QPs) /Bitrates. We use 6 QP points here to characterise the RD function for a clip. BDRATE for a particular clip is therefore used as a cost function in standard canned numerical routine (like Brent's [19] or Powell's algorithm [20]). The optimiser then simply iterates until changes in the cost function become small enough. See [5, 21, 22, 23] for more details.

Our previous work [5, 21] targeted VP9 and HEVC, treating all frames in the same way. Using the YouTube-UGC dataset [24] (~10k clips), we reported average BD-rate (%) gains of 1.63% for VP9 (libvpx-VP9) and 1.87% for HEVC (x265). Later in 2022 [22, 23], we recognised that most of the gains could be due to effects on keyframes and this effect was different for different codecs. By optimising $\lambda$ only in the higher priority frames (Keyframes, highest level B-frames, Golden-frames) we gained a 4-fold increase in BD-RATE for AV1: from 0.41% to 1.63%.

For this work, we deploy a similar strategy but optimise with respect to BD-rate (%) of Multi-scale Structural Similarity Index (MS-SSIM). Instead of a single multiplier, we estimate two different $k$ for two different keyframe types in AV1 seeking to improve gains further. Two different optimiser methods are used in this work, a) Brent's method for single $\lambda$ multiplier (x264, x265); and b) Powell's method for the 2-dimensional search for two $\lambda$ multipliers (different frame-types) for AV1 variants (libaom-AV1, SVT-AV1).

## Results

Details about the dataset and encoder settings used can be found in the Appendix. Table 1 summarizes the gains for optimised transcoding in our proposed process. Gains with this process are modest for AV1 (2%) but significant for selected clips, i.e., 10% of clips achieve gains of over 5% in AV1 for example, and 4% achieve similar gains in X264/5. Treating keyframes differently results in much improved average BDRATE gains with the per-clip gains of more than 10% for AV1. This shows that the per-clip approach can make a significant difference.

| *Video Encoder* | *Tuning Method* | *$k$* | *Average Iterations* | *BD-Rate (%) MS-SSIM Gains* | | | |
|---|---|---|---|---|---|---|---|
| | | | | *Average* | *Best* | *Clips with >1% gains* | *Clips with >5% gains* |
| *X264* | All frames | 1.20 | 9.8 | -1.54 | -9.89 | 40 | 4 |
| *X265* | All frames | 1.17 | 11.7 | -1.50 | -9.67 | 42 | 4 |
| *Libaom-av1* | K1 = KF; K2 = GF/ARF | (9.83, 1.70) | 48.27 | -2.49 | -14.89 | 68 | 12 |
| *SVT-AV1* | K1 = GF/ARF; K2 = Interframe | (1.44, 1.20) | 44.41 | -2.00 | -9.16 | 40 | 10 |

*TABLE 1: BDRATE GAINS WHEN OPTIMISING UGC-SUBSET (100 VIDEOS), FOR DIFFERENT ENCODERS WITH DIFFERENT TUNING METHODS.*

## Reducing Computational Costs

Clearly the main drawback of this kind of direct optimisation technique is that for each iteration of the optimiser, we require the evaluation of 6 RD points which implies 6 encodes. To make this process more practical we can instead perform the optimisation on a proxy of the original clip at a lower datarate or use a proxy encoder, e.g., the encoder at a faster preset. These ideas (applied to bitrate ladder design) were first presented by Katsavounidis et al at the Meta Industry workshop at ICIP in Taiwan 2019. In this case, the $k$ estimated from these proxies is then applied to the actual encoder/datarate clip. We expect to lose optimality at the higher resolution or actual encoder preset of course but choosing the datarate and/or faster preset judiciously can reduce this loss.

   a) **Method A: *Downsampling*.** We estimate $k$ at a lower resolution and apply it to the original resolution clip. For videos <=720p we recommend using a proxy at 144p. For videos above 720p, we recommend downsampling the spatial resolution by 2.
   b) **Method B: *Using a faster preset*.** We use the fastest preset for the encoder proxy (speed-6 for AV1)

Table 2 reports on results over our YT UGC set. We measure impact of the proxy strategies on reducing the compute time per iteration of the optimizer. There is a substantial speed-up with fractional loss in gains. For x264/x265, we deployed Method (a), and for AV1 variants (libaom-av1, svt-av1), we deployed Method (b). We achieved 80-100% of the possible BDRATE gains with compute time improvement of 20-200 in all cases except SVT-AV1, where we achieved only 50% of the possible gains with a x6 speed improvement, perhaps showing the efficiency of SVT-AV1. Given 40 optimisation iterations, the overall speedup for AV1 is a substantial 8000x!

| Video Encoder | Method | BD-rate MS-SSIM (%) | | Avg encoding time per optimisation iteration (mins) | |
| --- | --- | --- | --- | --- | --- |
| | | *Default* | *Proxy method* | *Default* | *Proxy method* |
| X264 | Downsample | -1.54 | -1.31 | 0.898 | 0.049 |
| X265 | Downsample | -1.50 | -1.23 | 3.246 | 0.148 |
| Libaom-AV1 | Proxy preset | -2.49 | -2.53 | 5817.10 | 25.28 |
| SVT-AV1 | Proxy preset | -2.00 | -0.87 | 27.73 | 5.487 |

TABLE 2: BD-RATE (%) MEASURED IN TERMS OF MS-SSIM BETWEEN DEFAULT AND PROXY-METHOD FOR OUR DATASET.

There is an additional overhead required to downsample the videos, an overhead that was excluded from the encoding time calculation. It is expected that other resolutions of video content would already be available in most content delivery systems. One explanation for the dramatic effect with libaom-av1 is due to the encoder preset configuration from AOM-CTC where we have controlled encoder settings in static GOP conditions.

**ML for Computational Efficiency**: Although DNNs are not discussed in the same room as low computational cost algorithms, the previous class of traditional ML techniques are quite computationally efficient. So, it is possible to propose that we can extract features from the original input clip and use those to predict an optimal $k$. Covell et al [25] were the first to try this approach; we use a feature set derived from their proposed set. It includes measures of complexity based on relative P/B frame bitrates measured from a first pass encode with default settings. See Appendix for details of the feature set. With this approach, we only need one RD operating point to be calculated, followed by assembly of the feature vector. We used a random forest regressor with squared error cost function to predict $k$. The computational complexity was therefore reduced by a factor of 60X. With the same dataset as used above, we achieved 67 % of the expected BDRATE gains with an average BD-rate improvement of 1.01% for H.265.

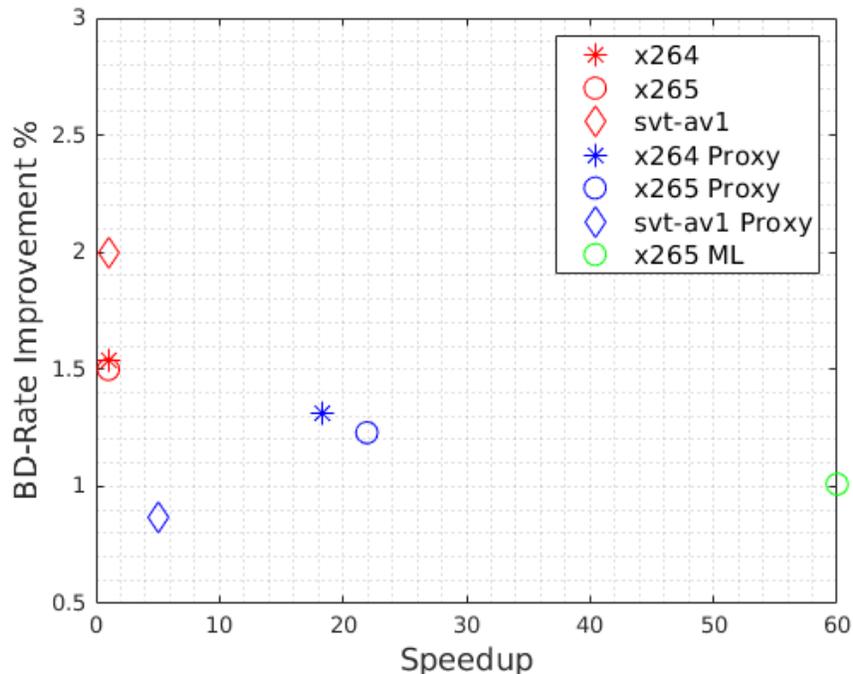

FIGURE 3: SPEEDUP IMPROVEMENT OF DIFFERENT METHODS. X-AXIS DENOTES SPEED-UP FACTOR OVER BASELINE; Y-AXIS DENOTES THE BD-RATE (%) IMPROVEMENT.

**Workflow:** An important point to note here is that we consider these optimisation modules to be best deployed as a 2$^{nd}$ pass action in a transcoding pipeline. That is because in all of these methods we always begin with a default encode pass from which inference begins. Hence we can easily track whether optimisation is worth it and terminate early on the basis of bitrate performance or computational load. Figure 3 shows the tradeoff between computational performance in terms of speed improvement and BD-Rate gains. As expected, computational load is proportional to bitrate gains. For x265, the ML technique is the fastest but worst performing, while the use of downsampled proxies perhaps yields the best tradeoff. Even though reduction in compute load is important, assessing the compute load ahead of time could also mitigate the pain of large compute loads from a user point of view. We discuss that functionality next.

## Section 3: Cloud cost calculator for budgeting cloud deployment

Current cloud based transcoding workflows facilitate predefined encoding settings e.g., AWS MediaConvert. These are billed according to a) framerate b) resolution c) encoder type d) standard encoding preset (Basic vs Professional Tier), and e) region of the cloud server. There is also reserved pricing for transcoding slots. For example, for a 10-minute HD 30 fps video, encoding with 2 different encoders (H.264 and H.265), in predefined AWS MediaConvert settings at US-East, will cost $0.15 for basic tier or $0.24/$0.42 (Speed/Quality variants) in professional tier. HEVC (only in professional tier) costs $0.48/$3.3 for the same options. Reserved pricing is $400 and upwards. *Note: These costs are measured at the time of writing this work in February 2023.* Custom encoder binaries (via FFmpeg

A paper from the *Proceedings of the 2023 NAB BEIT Conference*   7

software or standalone encoder binary) can also be deployed. In AWS, users can either use a custom AWS Lambda instance for jobs under 15 mins or deploy in an AWS Fargate instance for longer jobs.

The point is that against this backdrop of heterogeneous cost/compute profiles, given a large set of clips, it is unclear how a user can reliably assess the compute load required to execute the encoding jobs without performing the operation. A simple approach to cost prediction is to run a sample of the encoding jobs and measure the cost, hence predicting the overall cost by scaling by the number of clips. But that might incur unjustified costs itself. Since 2015 [26], there has been increasing attention to modelling and predicting encoding time for a given video based on the intrinsic properties of the clip. The essential idea is that the encoding time must be related in some way to the content of the clip and so features extracted from the clip should be able to help provide a good prediction. Initial proposals use video metadata as features [26] while later work ([27] circa 2020) used spatial and temporal statistical measures of clip complexity. The work by Amirpour et al [5] used a measure for spatial and temporal complexity derived from DCT coefficients (which we also use here) and an additional set of features generated by the MobileNet classification network [28]. Figure 4 shows the process flow which is common to most approaches to this problem so far. In our case the feature set consists of video properties like size, framerate together with statistical measures of spatial and temporal complexity and target encoder settings. The effect of training data on the performance of the final cost prediction almost swamps any effect from the particular classifier chosen.

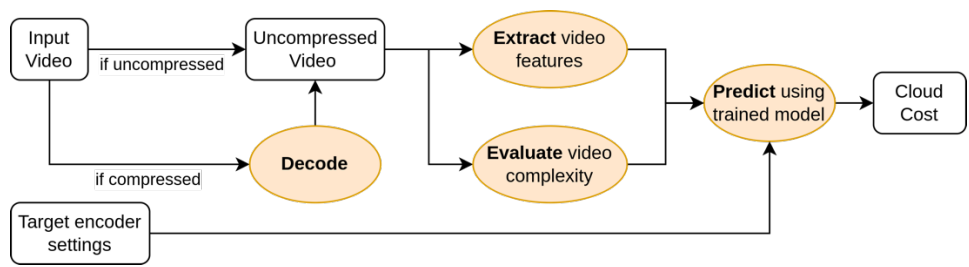

FIGURE 4: FLOW DIAGRAM OF A VIDEO-INSTRINSIC CLOUD COST CALCULATOR. FEATURE EXTRACTION ALONG WITH COMPLEXITY ESTIMATION IS CARRIED OUT FOR PREDICTION.

In previous efforts, the prediction algorithms directly predicted the transcoding time based on these features using regression approaches [27]. However, the datasets tend to be quite unbalanced with more data at the low compute load end rather than the high compute load end. The heavy tailed distribution of the resulting encoding times encourages the use of log scales for building regression models. Another approach to deal with this is to quantise clip encode durations in bins with increasing bin widths and use a classifier. A class then refers to a range of durations and not to a single execute time. We deployed a simple Support Vector Machine (SVM) classifier for this purpose and found that it improved recall by about 20% c.f. linear bin width. However other effects were more important and those are discussed in the next section in the context of a regressor. Our feature set is assembled from data extracted using the existing VideoComplexityAnalyzer (VCA) [29] toolkit. See the Appendix, Table A.4 for details. The compute time for executing this ML process is negligible w.r.t. to the time to encode.

**Analysis** Table 3 reports the overall performance of our system (H.264 over 6 CRFS 22:5:47) using a well known regression framework (XGBoost). Our various test setups are in Table A.5. The rows show performance of different test sets, while columns show the impact of linear and log scales for time measurement. R2 scores show that using the log space improves fitting across all the sets. From a customer standpoint however, it is more interesting to report on % error in the estimation of the actual encode time. We use the MAE and symmetric MAE (sMAE) which is less biased to errors in small numbers. On the face of it, we appear to be able to achieve an error of 5-13% with this regressor on our training set using the log-scale. The "Generalised" row shows the performance when we ensure that the content sources in train/test are different. That is the usual approach to generating models which generalise well. However, from a cost budget standpoint the linear scale is more appropriate. Measuring that error in linear space, the sMAE is about 12% (Generalised) and MAE at 25%. So, in fact whether we use linear or log time transformations we are over 10% in error.

|  |  | Linear |  |  | Log |  |  | Log →Linear |  |
|---|---|---|---|---|---|---|---|---|---|
| Test Set | Split Method | R2 | MAE (%) | sMAE (%) | R2 | MAE (%) | sMAE (%) | MAE (%) | sMAE (%) |
| Validation Set | Overfit | 0.987 | 40.09 | 13.09 | 0.998 | 13.83 | 5.11 | 10.25 | 5.04 |
| Validation Set | Generalised | 0.762 | 98.22 | 25.43 | 0.969 | 26.92 | 8.69 | 24.56 | 12.05 |
| YT-UGC Set | Overfit | -0.058 | 225.41 | 40.65 | 0.847 | 65.82 | 18.01 | 44.67 | 17.45 |
| YT-UGC Set | Generalised | -0.074 | 181.79 | 35.79 | 0.818 | 115.95 | 16.99 | 54.77 | 18.90 |

TABLE 3: PERFORMANCE OF VARIOUS ML ALGORITHMS TRAINED WITH DIFFERENT DATASETS INCLUDING TRANSFORMATION OF TIME TO LINEAR SPACE.

With an error rate potentially as high as 25% the customer can be confident only of estimates in orders of magnitude. For example, if the cost of the actual encode was 100 cycles, we would be able to confidently predict the range was 75 – 125 cycles. At a cost of 10,000 cycles our prediction range increases to 7,500-12,500. The effect in real terms of cycles is therefore exaggerated for more compute intensive jobs. This is not necessarily as poor as the numbers might imply. Whereas without this tool, the customer had no information, now we can be confident of some range based on the actual data.

To show how important the influence of content is on the performance of these models, we change our training set so that clips from the same long form content appear both in the holdout and training set. This is the "Overfit" row. In that case we can achieve 5% (log) or 13% (linear) sMAE in training and test. By including samples of the target data in the training set, we have reduced the error from about 25% to about 10%. That means prediction is significantly improved. Our 10,000 cycle count can now be predicted as within a range of 9000 – 11,000 cycles. This is a well-known phenomenon in ML, but in this application, overfitting does have a use case. When presented with a large corpus, it means that it is possible for a small training set created using a random selection of clips from that corpus, can yield good prediction performance. That is a reliable workflow.

When we use this model on a new corpus which was not part of the training/testing (YT-UGC), performance degrades across the board. In this case "Overfit" and "Generalised" just refers to the model trained on data exhibiting "crosstalk between train/test" or not accordingly. In that case error rates increase dramatically. Further insight about the nature of this holdout test set as compared to the training data is shown in the distribution of Spatial and Temporal Energy (SE/TE) for the clips in Figure 5. The YT-UGC set seems to be approximately distributed in the same space as the training set but not at the higher end of the scale, yet the prediction error is higher than we would expect. This means that SI/TI may not correlate well enough to encode time and that more useful intrinsic information needs to be extracted. Essentially, the ability of these ML approaches to generalise remains a question. For example, by retraining different models for each preset/target CRF combination (not shown in table) for 4K resolution videos, we can reduce the final MAE (%) to as low as 3.8% (with CRF27, and very-slow preset). This also indicates a significant body of work left to be done to determine how best to sample this space.

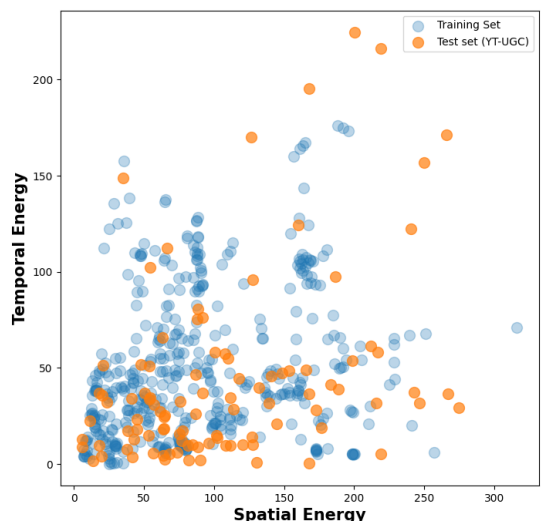

FIGURE 5: DISTRIBUTION OF SPATIAL AND TEMPORAL ENERGY FOR OUR TRAINING AND TEST CLIPS MEASURED USING VCA [24].

## Conclusion

In this work, we examined 3 tools addressing gaps in the development of transcoder pipelines: optimisation for pre-processing, encoder parameterisation, and a tool for compute load prediction. We also proposed low-cost solutions in each case which are appropriate for cloud deployment. Our use of proxies was able to reduce the compute load for RD optimisation by a factor of 200 without loss of quality. The pre-processing module relies only on a simple 2D polynomial fit to data that can be gathered offline. Finally, our cloud cost calculator is more reliable in an overfitting mode than in a general "blind" mode. There is still substantial room for improvement in these tools, and we expect that the class of ML cloud cost predictors has enormous potential for customer management.

## Acknowledgement

This work was supported by Enterprise Ireland's (EI) Disruptive Technology Innovation Fund (DTIF) with Grant Number DT-2019-0068 (VISP), The Adapt SFI Research Centre SFI Grant Number (13/RC/2106), Google/YouTube Faculty Awards and the Ussher Research Studentship from Trinity College Dublin.

## References

[1] Aaron, A., Li, Z., Manohara, M., De Cock, J., & Ronca, D. (2015). "Per-title encode optimization." *The Netflix Techblog*.

[2] Reznik Y., J. Cenzano and B. Zhang, "Transitioning Broadcast to Cloud," in SMPTE Motion Imaging Journal, vol. 130, no. 9, pp. 18-32, Oct. 2021, doi: 10.5594/JMI.2021.3106162.

[3] van Roosmalen, P. M., Kokaram, A. C., & Biemond, J. (1998, September). "Noise reduction of image sequences as preprocessing for MPEG2 encoding." In *9th European Signal Processing Conference (EUSIPCO 1998)* (pp. 1-4). IEEE.

[4] Segall, C. A., Karunaratne, P. V., & Katsaggelos, A. K. (2000, December). "Preprocessing of compressed digital video." In *Visual Communications and Image Processing 2001* (Vol. 4310, pp. 163-174). SPIE.

[5] Ringis, D., Pitié, F., & Kokaram, A. (2020). "Per Clip Lagrangian Multiplier Optimisation for HEVC." *IS&T International Symposium on Electronic Imaging Science and Technology*, *32*(10), 136–137.

[6] Amirpour, H., Rajendran, P. T., Menon, V. V., Ghanbari, M., & Timmerer, C. (2022, September). "Light-weight Video Encoding Complexity Prediction using Spatio Temporal Features." In *2022 IEEE 24th International Workshop on Multimedia Signal Processing (MMSP)* (pp. 1-6). IEEE.

[7] Guleryuz, O. G., Chou, P. A., Hoppe, H., Tang, D., Du, R., Davidson, P., & Fanello, S. (2022, December). Sandwiched Image Compression: Increasing the resolution and dynamic range of standard codecs. In 2022 Picture Coding Symposium (PCS) (pp. 175-179). IEEE.

[8] Chadha, A., & Andreopoulos, Y. (2021). Deep perceptual preprocessing for video coding. In Proceedings of the IEEE/CVF Conference on Computer Vision and Pattern Recognition (pp. 14852-14861).

[9] Chadha, A., Anam, M. A., Treder, M., Fadeev, I., & Andreopoulos, Y. (2022). Toward Generalized Psychovisual Preprocessing For Video Encoding. SMPTE Motion Imaging Journal, 131(4), 39-44.

[10] Andreopoulos, Y., "Neural Pre and Post-Processing for Video Encoding With AVC, VP9, and AV1," in AOM Research Symp, 2022.

[11] Wang, H., Gan, W., Hu, S., Lin, J. Y., Jin, L., Song, L., ... & Kuo, C. C. J. (2016, September). "MCL-JCV: a JND-based H. 264/AVC video quality assessment dataset." In *2016 IEEE international conference on image processing (ICIP)* (pp. 1509-1513). IEEE.

[12] Kokaram, A. (1994). 3D Wiener filtering for noise suppression in motion picture sequences using overlapped processing. *Signal Processing VII, Vol3*, 1780-1783.

[13] Tassano, M., Delon, J., & Veit, T. (2019, September). "Dvdnet: A fast network for deep video denoising ." In *2019 IEEE International Conference on Image Processing (ICIP)* (pp. 1805-1809). IEEE.

[14] Hanooman, V., Kokaram, A. C., Su, Y., Birkbeck, N., & Adsumili, B. (2022, October). "An Empirical Approach for Optimising the Impact of a Preprocessor in a Transcoding Pipeline." In *2022 IEEE International Conference on Image Processing (ICIP)* (pp. 2201-2205). IEEE.

[15] Katsavounidis, I., & Guo, L. (2018, September). "Video codec comparison using the dynamic optimizer framework." In *Applications of Digital Image Processing XLI* (Vol. 10752, pp. 266-281). SPIE.

[16] Sullivan, G. J., & Wiegand, T. (1998). "Rate-distortion optimization for video compression." *IEEE signal processing magazine*, *15*(6), 74-90.

[17] Zhang, F., & Bull, D. R. (2018). "Rate-distortion optimization using adaptive lagrange multipliers." *IEEE Transactions on Circuits and Systems for Video Technology*, *29*(10), 3121-3131.

[18] Bjontegaard, G. (2001). "Calculation of average PSNR differences between RD-curves." *ITU SG16 Doc. VCEG-M33*.

[19] Brent, R. P. (2013). *Algorithms for minimization without derivatives*. Courier Corporation.

[20] Powell, M. J. (1964). An efficient method for finding the minimum of a function of several variables without calculating derivatives. *The computer journal*, 7(2), 155-162.

[21] Ringis, D. J., Pitié, F., & Kokaram, A. (2020, August). "Per-clip adaptive lagrangian multiplier optimisation with low-resolution proxies." In *Applications of Digital Image Processing XLIII* (Vol. 11510, pp. 40-51). SPIE.

[22] Vibhoothi, Pitié, F., Katsenou, A., Ringis, D. J., Su, Y., Birkbeck, N., Lin, J., Adsumilli, B. & Kokaram, A. (2022, October). "Direct optimisation of λ for HDR content adaptive transcoding in AV1." In *Applications of Digital Image Processing XLV* (Vol. 12226, pp. 36-45). SPIE.

[23] Vibhoothi, Pitié, F., & Kokaram, A. (2022, October). "Frame-Type Sensitive RDO Control for Content-Adaptive Encoding." In *2022 IEEE International Conference on Image Processing (ICIP)* (pp. 1506-1510). IEEE.

[24] Wang, Y., Inguva, S., & Adsumilli, B. (2019, September). "YouTube UGC dataset for video compression research." In *2019 IEEE 21st International Workshop on Multimedia Signal Processing (MMSP)* (pp. 1-5). IEEE.

[25] Covell, M., Arjovsky, M., Lin, Y. C., & Kokaram, A. (2016). "Optimizing transcoder quality targets using a neural network with an embedded bitrate model." *Electronic Imaging, 2016*(2), 1-7.

[26] Deneke, T., Lafond, S., & Lilius, J. (2015, December). "Analysis and transcoding time prediction of online videos." In *2015 IEEE International Symposium on Multimedia (ISM)* (pp. 319-322).

[27] Zabrovskiy, A., Agrawal, P., Mathá, R., Timmerer, C., and Prodan, R. Complexcttp: Complexity class based transcoding time prediction for video sequences using artificial neural network. In 2020 IEEE Sixth International Conference on Multimedia Big Data (BigMM) (2020), pp. 316–325

[28] Howard, A. G., Zhu, M., Chen, B., Kalenichenko, D., Wang, W., Weyand, T., Andreetto, M., & Adam, H. (2017). "Mobilenets: Efficient convolutional neural networks for mobile vision applications". *arXiv preprint arXiv:1704.04861*.

[29] Menon, V. V., Feldmann, C., Amirpour, H., Ghanbari, M., & Timmerer, C. (2022, June). VCA: Video complexity analyzer. In *Proceedings of the 13th ACM Multimedia Systems Conference* (pp. 259-264).

# Appendix

## Experimental Setup

We demonstrate performance with 4 different implementations of three coding standards: x264 (H.264), x264 (H.265), libaom-av1 and SVT-AV1 (AV1). More information on the configuration is shown in Table A.1.

| Video Encoder | Version | QP | Tuning used | Encoder Configuration |
|---|---|---|---|---|
| **x264** | 0.157.x | {22, 27, 32, 37, 42} | All frames | CRF mode single threaded, single pass, medium preset (default) <br> *$x264 –threads 1 --crf $QP --output OUT.264 $INPUT.Y4M* |
| **x265** | 3.0+28-gbc05b8a91 | {22, 27, 32, 37, 42} | All frames | CRF mode single threaded, single pass, medium preset (default) <br> *$x265 –threads 1 --input INPUT.y4m --crf QP   --output OUT.mp4$* |
| **libaom-av1** | 3.2.0, 287164d | {27, 39, 49, 59, 63} | K1 = KF; K2 = GF/ARF | AOM-CTC Random Access@P6, Static GOP, Single-threaded. <br> *$AOMEMC --cpu-used=0 --passes=1 --lag-in-frames=19 --auto-alt-ref=1 --min-gf-interval=16 --max-gf-interval=16 --gf-min-pyr-height=4 --gf-max-pyr-height=4 --limit=130 --kf-min-dist=65 --kf-max-dist=65 --use-fixed-qp-offsets=1 --deltaq-mode=0 --enable-tpl-model=0 --end-usage=q --cq-level=$QP --enable-keyframe-filtering=0 --threads=1 --test-decode=fatal -o output.ivf $OPTIONS $INPUT.Y4M* |
| **SVT-AV1** | v1.4.1-17, gf0efc5fa | {27, 33, 39, 46, 52, 58}, | K1 = GF/ARF; K2 = InterFrames | CRF mode with 1 LogicalProcessor@P9 <br> *$SVTAV1ENC -lp 1 --crf $QP --preset 9 -I $INPUT.Y4M -o output.ivf* |

*TABLE A.1: VIDEO ENCODER CONFIGURATIONS, QP SETTINGS, AND SAMPLE COMMAND-LINE OPTIONS*

## Dataset

We implemented our direct search optimization scheme on one hundred clips from the YouTube-UGC [24] dataset. These videos were selected to provide a robust sample of modern video. This list can be found in Table A.2.

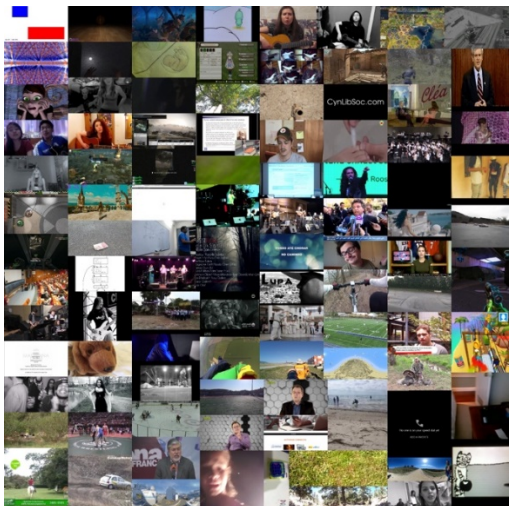

FIGURE A.1: SCREENSHOT OF THE UGC-SUBSET1 DATASET CONTAINING 100 VIDEOS USED IN THE STUDY.

| # | Video Group | Number of videos | Video count (resolution, count) |
|---|---|---|---|
| 1 | HDR | 2 | {(2160, 1), (1080, 1)} |
| 2 | HowTo | 4 | {(720, 1), (360, 1), (1080, 1), (480, 1)} |
| 3 | VR | 4 | {(720, 1), (2160, 1), (960, 1), (1080, 1)} |
| 4 | LyricVideo | 4 | {(720, 1), (360, 1), (1080, 1), (480, 1)} |
| 5 | NewsClip | 4 | {(720, 1), (360, 1), (1080, 1), (480, 1)} |
| 6 | Vlog | 5 | {(720, 1), (360, 1), (480, 1), (2160, 1), (1080, 1)} |
| 7 | VerticalVideo | 7 | {(720, 1), (1280, 1), (360, 1), (480, 1), (2160, 1), (960, 1), (1080, 1)} |
| 8 | CoverSong | 8 | {(1080, 2), (480, 2), (720, 2), (360, 2)} |
| 9 | Animation | 8 | {(1080, 2), (480, 2), (720, 2), (360, 2)} |
| 10 | Lecture | 8 | {(1080, 2), (480, 2), (720, 2), (360, 2)} |
| 11 | MusicVideo | 8 | {(1080, 2), (480, 2), (720, 2), (360, 2)} |
| 12 | TelevisionClip | 8 | {(1080, 2), (480, 2), (720, 2), (360, 2)} |
| 13 | LiveMusic | 8 | {(1080, 2), (480, 2), (720, 2), (360, 2)} |
| 14 | Gaming | 10 | {(1080, 2), (480, 2), (2160, 2), (720, 2), (360, 2)} |
| 15 | Sports | 12 | {(1080, 2), (480, 2), (2160, 2), (960, 2), (720, 2), (360, 2)} |
| **Total** | | **100** | **{360, 21}, {480, 21}, {720, 22}, {960, 4}, {1080, 23}, {1280, 1}, {2160, 8}** |

TABLE A.2: TESTING DATASET OF 100 VIDEOS FROM THE YOUTUBE-UGC DATASET FROM 15 DIFFERENT CLASSES CONSISTING OF 7 DIFFERENT SPATIAL RESOLUTIONS.

## Feature set for prediction of k

Features from the encoded video were used to predict the optimal k for improved BD-Rate performance for a given clip in x265. These features were collected from encoding the clips at QP = 32 when k = 1. Further details about the model can be found in [21].

| YouTube UGC Group | P avg-QP | P/B Count | P/B Ratio Y *P/B Size |
|---|---|---|---|
| WIDTH | P bitrate | P/B bitrate | P/B Ratio U *P/B Size |
| HEIGHT | B count | bitrate *P/B Ratio Y | P/B Ratio V *P/B Size |
| Bitrate | B avg-QP | bitrate *P/B Ratio U | P/B Count *P/B Size |
| Y-MS-SSIM (for each frame type) | B bitrate | bitrate *P/B Ratio V | P/B Ratio Y *P/B Ratio U |
| U-MS-SSIM (for each frame type) | P/B Ratio Y | bitrate *P/B Count | P/B Ratio Y *P/B Ratio V |
| V-MS-SSIM (for each frame type) | P/B Ratio U | bitrate *P/B Size | P/B Ratio U *P/B Ratio V |
| P count | P/B Ratio V | | |

*TABLE A.3: FEATURES USED TO TRAIN RANDOM FOREST MODEL USED TO PREDICT K FOR LAGRANGIAN MULTIPLIER OPTIMISATION.*

## Feature set for prediction of encoding time

| Features |
|---|
| Height |
| Number of total pixels |
| Frame rate |
| Number of frames |
| Spatial Energy [Mean, Max, Median, Std deviation] |
| Temporal Energy [Mean, Max, Median, Std deviation] |
| Mean Brightness |
| Preset |
| Target CRF |

*TABLE A.4: FEATURES USED TO TRAIN ENCODING TIME PREDICTOR*

## Experimental setup for encoding time prediction

| | |
|---|---|
| **Dataset used** | AOM-CTC, AOM-Candidates, Vimeo-Corpus-10s, Xiph-Media Collection (commonly known as *Derf* video set). |
| **Number of videos** | 451 |
| **Resolution** | 4096x2160, 3840x2160: 4K (107); 1920x1080, 1080x1920: 2K (182); 1280x720: 720p (75) 480x360: 360p (66), 480x270: 270p (21) |
| **Duration** | 2s, 4s |
| **CRF** | 22, 27, 32, 37, 42, 47 |
| **Video preset** | ultrafast, superfast, veryfast, faster, fast, medium, slow, slower, veryslow |

*TABLE A.5: EXPERIMENTAL SETUP FOR ENCODING TIME PREDICTOR*